\documentclass[aip,cha, reprint,showpacs,showkeys]{revtex4-1}
\usepackage{amssymb, dcolumn,  graphicx, latexsym, amsfonts, bm}

\begin{document}
 
\title{The ion-acoustic instability of the inductively coupled plasma driven by the ponderomotive electron current formed in the skin layer}
\author{V. V. Mikhailenko}\email[E-mail: ]{vladimir@pusan.ac.kr}
\affiliation{Plasma Research Center, Pusan National University,  Busan 46241, South Korea.}
\affiliation{BK21 Plus Information Technology, Pusan National University,  Busan 46241, South Korea.}
\author{V. S. Mikhailenko}\email[E-mail: ]{vsmikhailenko@pusan.ac.kr}
\affiliation{Plasma Research Center, Pusan National University,  Busan 46241, South Korea.}
\author{Hae June Lee}\email[E-mail: ]{haejune@pusan.ac.kr}
\affiliation{Department of Electrical Engineering, Pusan National University, Busan 46241, South Korea.}

\begin{abstract}
The stability theory of the inductively coupled plasma (ICP) is developed for the case when the electron 
quiver velocity in RF wave is of the order of or is larger than the electron thermal velocity. The theory predicts the existence 
the instabilities of the ICP which are driven by the current formed in the skin layer by the accelerated  electrons, which move relative 
 ions under the action of the ponderomotive force. 
\end{abstract}
%\pacs{52.35.Qz}

%\noindent{\it Keywords\/ kinetic theory, drift Alfven instability, shear flow}

\maketitle

\section{Introduction}\label{sec1}
The regimes of the anomalous skin effect (or nonlocal regime)\cite{Weibel, Kolobov} is typical for the low pressure inductive plasma sources 
employed in material processes applications\cite{Lieberman}. It occurs when the frequency $\omega_{0}$ of the operating 
electromagnetic (EM) wave is much above the electron-neutral collision frequency, but much less than the electron plasma frequency. 
In this regime, the interaction of the EM field with electrons is governed by the electron thermal motion. For this reason, the EM wave absorption
\cite{Shaing}, the formation of the anomalous skin layer near the plasma boundary\cite{Weibel}, and the anomalous 
electron heating\cite{Aliev, Tyshetskiy} require the kinetic description which involves the well known mechanism of collisionless power dissipation - Landau 
damping. It stems from the resonant wave-electron interaction under condition that the electron thermal velocity $v_{Te}$ is comparable with (or is larger 
) the EM phase velocity. The theory of the anomalous skin effect is developed as a rule, employing the linear approximation to the solution of the 
Vlasov equation for the electron distribution function. It is assumed in this theory that the equilibrium electron distribution function depends only 
on the electron kinetic energy and does not involve the electron motion in the time dependent spatially inhomogeneous EM wave. 
This approximation is valid when the quiver velocity of electron in the EM wave is negligible in comparison with the electron thermal velocity.

It was found experimentally \cite{Godyak1, Godyak2, Godyak3, Godyak4} and analytically\cite{Cohen1,Cohen2,Smolyakov1,Piejak,Smolyakov2} 
that at the low driving frequency of an inductive discharge, at which RF Lorentz force acting on electrons becomes comparable 
to or larger than the RF electric field force, the nonlinear effects in the skin layer becomes essential. The theoretical analysis 
\cite{Cohen1,Cohen2, Smolyakov1,Piejak,Smolyakov2, Froese} has shown that electron oscillatory motion in the inhomogeneous RF field in the skin layer leads 
to the ponderomotive force. This force is regarded as the responsible for the reduction of the steady state electron density distribution within the 
skin layer\cite{Cohen1,Cohen2} and  for the formation of experimentally observed\cite{Godyak1, Godyak2, Godyak3, Godyak4} and analytically predicted
\cite{Smolyakov1,Piejak,Smolyakov2} second harmonics which was found \cite{Godyak1, Godyak2, Godyak3, Godyak4} to be much larger than the electric field on 
the fundamental frequency. It was found that the lowfrequency/high-amplitude portion of the anomalous skin effect
regime changes behaviour so completely that it was systematized as the nonlinear skin effect regime\cite{Froese}.

Our paper is devoted to the analytical investigations of the nonlinear processes in the high frequency range\cite{Kolobov, Alexandrov} of the driven 
frequency $\omega_{0}$,
\begin{eqnarray}
&\displaystyle
\omega_{pe}\frac{v_{Te}}{c}\ll \omega_{0}\ll \omega_{pe},
\label{1}
\end{eqnarray}
where $\omega_{pe}$ is the electron plasma frequency and $v_{Te}$ is the electron thermal velocity. The skin effect in this frequency range is determined as 
the classical or normal skin effect for which the collisionless skin depth  $L_{s}$ is equal to $c/\omega_{pe}$ \cite{Kolobov, Alexandrov}. 
The linear theory of the classical skin effect predicts\cite{Kolobov}, that the wave is reflected from the plasma without energy dissipation 
in the skin layer. This result was derived under 
the assumption of the cold plasma which consists in neglecting the thermal motion of electrons. In this case, a situation can occur 
that the electron quiver 
velocity in a skin layer under the action of the EM wave approaches or is larger than the electron thermal velocity 
and the EM electric field force acting on 
electrons prevails over the Lorentz force formed by the magnetic field of the EM wave. 
Under such conditions it is reasonable to talk about free oscillations of a plasma particle 
under the action of the EM field (at least, in the zero approximation in the ratio of the 
collision frequency to the field frequency). The relative oscillatory motion of the electrons and ions in the RF field is a potential 
source of numerous instabilities of the parametric type (see, for example, Refs.\cite{Silin,Porkolab,Mikhailenko1,Akhiezer}) 
with frequencies $\omega$  comparable with or 
less than the frequency $\omega_{0}$ of the applied RF wave. It is clear that in such a situation  an essentially 
nonlinear dependence of the plasma conductivity on the RF field as well as 
the anomalous absorption of the RF energy due to the development of the plasma 
turbulence and turbulent scattering of electrons arise\cite{Silin,Porkolab,Mikhailenko1,Akhiezer}. 
This is just the case which interest us in the present paper.

It is usually accepted in the theoretical investigations of the parametric instabilities 
excited by the strong EM wave in the unbounded uniform plasmas, 
that the approximation of the spatially homogeneous pump wave may suffice since the parametrically 
excited waves have the wave number much larger than the wave number of the pump wave. The 
presence of the skin layer at the plasma boundary near the RF 
antenna, in which the amplitude of the EM wave attenuates, requires the development of new 
approach to the theory of the instabilities of the parametric 
type in  which the spatial inhomogeneity of the pumping wave is accounted for. This new 
kinetic approach, grounded on the methodology of the oscillating 
modes, is developed in Sec. \ref{sec2}. We found that in the skin layer electrons experience 
the oscillating motion in RF field jointly with the uniformly accelerated motion under the action of the 
ponderomotive force resulted from the spatial 
inhomogeneity of the RF field in this layer. The basic equation for the perturbed 
electrostatic potential which determines the stability of the inductively 
coupled plasma against the development of the electrostatic instabilities in skin layer is 
derived in Sec. \ref{sec3}. We found, that the accelerated motion of the electrons in the skin layer 
is the dominant factor in the instabilities development. The linear theory of the instabilities, driven by the 
accelerated motion of electrons relative to ions under the action of the ponderomotive force was developed in Sec. \ref{sec4}.
Conclusions are presented in Sec. \ref{sec5}. 

\section{Basic transformations and governing equations}\label{sec2}
We consider a model of a plasma occupying region $z\geqslant 0$. The RF antenna which launches the RF wave with frequency $
\omega_{0}$ is assumed to exist to the left of the plasma boundary $z=0$. The electric, $
\mathbf{E}_{0}\left(z,t\right)$, and magnetic, $\mathbf{B}_{0}\left(z,t\right)$, fields of a 
such RF wave, are directed along the plasma boundary and attenuate along $z$ due to the skin effect. In the frequency range 
(\ref{1}), these fields are exponentially decaying with $z$, and are sinusoidally varying with time, 
\begin{eqnarray}
&\displaystyle 
\mathbf{E}_{0}\left(z,t\right)=E_{0y}e^{-\kappa z}\sin \omega_{0}t\mathbf{e}_{y},
\label{2}
\end{eqnarray}
and 
\begin{eqnarray}
&\displaystyle 
\mathbf{B}_{0}\left(z,t\right)=E_{0y}\frac{c\kappa}{\omega_{0}}e^{-\kappa z}\cos \omega_{0}t \mathbf{e}_{x},
\label{3}
\end{eqnarray}
where $\mathbf{E}_{0}$ and $\mathbf{B}_{0}$ satisfy the Faraday’s law, $\partial E_{0}/
\partial z=\partial B_{0}/c\partial t$, and   $\kappa^{-1}=L_{s}$
is the skin depth for the classical skin effect\cite{Kolobov, Alexandrov}, 
\begin{eqnarray}
&\displaystyle L_{s}=\frac{c}{\omega_{pe}}.
\label{4}
\end{eqnarray}
In this paper, we consider the effect of the relative motion of plasma species in the applied RF field on the 
development the short scale electrostatic perturbations in the skin layer with wavelength much less than the skin layer depth. 
Our theory bases on the Vlasov equation for the velocity distribution function $F_{\alpha}$ of $\alpha$ species 
($\alpha=e$ for electrons and $\alpha=i$ for ions), 
\begin{eqnarray}
&\displaystyle 
\frac{\partial F_{\alpha}}{\partial t}+\mathbf{v}\frac{\partial F_{\alpha}}
{\partial\mathbf{r}}+\frac{e_{\alpha}}{m_{\alpha}}\left(\mathbf{E}_{0}
\left(z,t\right)+\frac{1}{c}\left[\mathbf{v}\times\mathbf{B}_{0}\left(z,t \right)\right] \right.
\nonumber
\\ 
&\displaystyle
-\nabla \varphi\left(\mathbf{r},t\right) 
\left)\frac{\partial F_{\alpha}}{\partial\mathbf{v}}\right. =0.
\label{5}
\end{eqnarray}
This equation contains the potential $\varphi\left(\mathbf{r},t\right)$ of the electrostatic plasma perturbations which is 
determined by the Poisson equation
\begin{eqnarray}
&\displaystyle \vartriangle \varphi\left(\mathbf{r},t\right)=
-4\pi\sum_{\alpha=i,e} e_{\alpha}\int f_{\alpha}\left(\mathbf{v},
\mathbf{r}, t \right)d\textbf {v}_{\alpha}, \label{6}
\end{eqnarray}
where  $f_{\alpha}$ is the perturbation of the equilibrium distribution function $F_{0\alpha}$,
$F_{\alpha}=F_{0\alpha}+f_{\alpha}$. The equilibrium distribution function $F_{0\alpha}$ is a function of the canonic momentums $p_{z}=m_{\alpha}v_{z}$ and 
$p_{y}=m_{\alpha}v_{y}-\frac{e}{c}A_{0y}\left(z, t\right)$, which are the integrals of the Vlasov equation (\ref{5}) without potential $\varphi
\left(\mathbf{r},t\right)$. It will be assumed to have a form
\begin{eqnarray}
&\displaystyle 
F_{\alpha 0}\left(v_{y},v_{z},z,t \right) =\frac{n_{0\alpha}}{2\pi v^{2}_{T\alpha} }
\exp\left[-\frac{v^{2}_{z}}{2v^{2}_{T\alpha}}\right.
\nonumber
\\ 
&\displaystyle
\left.-\frac{1}{2v^{2}_{T\alpha}}\left(v_{y}
-\frac{e}{cm_{\alpha}}A_{0y}\left(z,t \right) \right)^{2} \right],
\label{7}
\end{eqnarray}
where $n_{0\alpha}$ and $v_{T\alpha}=\left(T_{\alpha}/m_{\alpha}\right)^{1/2}$ are the equilibrium density and the thermal velocity of the $\alpha$ species 
particles with temperature $T_\alpha$, respectively. 
The electromagnetic potential $A_{0y}\left(z,t \right)$ for the electromagnetic field (\ref{2}) and (\ref{3}) is equal to\cite{Cohen1, Cohen2} 
\begin{eqnarray}
&\displaystyle 
\mathbf{A}_{0y}\left(z,t \right)= \frac{c\mathbf{E}_{0y}}{\omega_{0}}e^{-\kappa z}\cos \omega_{0}t.
\label{8}
\end{eqnarray}

The kinetic theory of the plasma stability with the time dependence of the $F_{\alpha 0}$ caused by 
the strong spatially homogeneous oscillating electric field $\mathbf{E}_{0}\left(t\right)=E_{0y}
\sin \omega_{0}t\mathbf{e}_{y}$ was developed\cite{Silin, Porkolab}  by employing the transformation $\mathbf{v}=\mathbf{v}_{\alpha}
+\mathbf{V}_{\alpha 0}\left(t\right)$ of the velocity variable $\mathbf{v}$ in the Vlasov equations for ions and electrons to velocity 
$\mathbf{v}_{\alpha}$ determined in the frame of references which oscillates with velocity $\mathbf{V}_{\alpha 0}\left(t\right)$ of particles 
of species $\alpha$  in velocity space, leaving unchanged position coordinates. With new velocity $\mathbf{v}_{\alpha}$ the explicit time 
dependence which stems from  the RF field is excluded from the Vlasov equation. In this paper we employ more general transformation of the 
velocity and position coordinates to the convected-oscillating frame of references determined by the relations
\begin{eqnarray}
&\displaystyle 
\mathbf{v}_{\alpha}=\mathbf{v}-\mathbf{V}_{\alpha}\left(\mathbf{r},t \right) , 
\nonumber
\\ 
&\displaystyle
\mathbf{r}_{\alpha}=\mathbf{r}-\mathbf{R}_{\alpha}\left(\mathbf{r},t \right)
= \mathbf{r} -\int\limits^{t} \mathbf{V}_{\alpha}\left(\mathbf{r},t_{1} \right)dt_{1}.
\label{9}
\end{eqnarray}
This transformation was decisive in the development of the parametric weak turbulence theory
\cite{Mikhailenko1}, and the theory of the stability and turbulence of plasma in pumping wave with finite wavelength\cite{Mikhailenko2} and admits the 
solution of the Vlasov equation in the case of the oscillating spatially inhomogeneous RF field. The transformation of Eq. (\ref{5}) for $F_{e}$
to velocity $\mathbf{v}_{e}$ and coordinate $\mathbf{r}_{e}$ variables determined by Eq. (\ref{9}) transforms Eq. (\ref{5}) to the form
\begin{eqnarray}
&\displaystyle 
\frac{\partial F_{e}\left(\mathbf{v}_{e},\mathbf{r}_{e},t \right) }{\partial t}
+\mathbf{v}_{e}\frac{\partial F_{e}}{\partial\mathbf{r}_{e}}
-v_{ej}\int\limits^{t}_{t_{0}}\frac{\partial V_{ek}\left( \mathbf{r},t_{1}\right) }{\partial r_{j}}dt_{1}
\frac{\partial F_{e}}{\partial r_{ek}}
\nonumber
\\ 
&\displaystyle
-v_{ej}\frac{\partial V_{ek}\left( \mathbf{r}_{e},t\right)}{\partial r_{ej}}\frac{\partial F_{e}}{\partial v_{ek}}-V_{ej}\left( \mathbf{r}_{e},t\right)
\int\limits^{t}_{t_{0}}\frac{\partial V_{ek}\left( \mathbf{r},t_{1}\right) }{\partial r_{j}}dt_{1}\frac{\partial F_{e}}{\partial r_{ek}}
\nonumber
\\ 
&\displaystyle
+\frac{e}{m_{e}}\left(\nabla \varphi\left( \mathbf{r},t\right)
-\frac{1}{c}\Big[\mathbf{v}_{e}\times\mathbf{B}
_{0}\left(z,t \right) \Big]\right)\frac{\partial F_{e}}{\partial\mathbf{v}_{e}}
\nonumber
\\ 
&\displaystyle
-\left\{
\frac{\partial V_{ej}\left( \mathbf{r},t\right) }{\partial t}+
V_{ek}\left( \mathbf{r},t\right)\frac{\partial V_{ej}\left( \mathbf{r},t\right) }{\partial r_{k}}
\right.  
\nonumber
\\ 
&\displaystyle
\left. 
+ \frac{e}{m_{e}}\left(\mathbf{E}_{0y}\left(z,t \right) 
+\frac{1}{c}\Big[\mathbf{V}_{e}\left( \mathbf{r},t\right)\times\mathbf{B}_{0}\left(z,t \right) \Big]\right)_{j} 
\right\}
\nonumber
\\ 
&\displaystyle
\times\frac{\partial F_{e}\left(\mathbf{v}_{e},\mathbf{r}_{e},t \right)}{\partial v_{ej}}=0.
\label{10}
\end{eqnarray}
In the approximation of the spatially uniform RF field (i. e. for $\kappa =0$ in our case), the time dependent RF electric field is excluded from Eq. 
(\ref{10}) for the velocity  $\mathbf{V}_{e}$ for which the expression in braces vanishes. In the case of the spatially inhomogeneous RF fields, this
selection of the velocity $\mathbf{V}_{e}$ provides the derivation of the solution for $F_{e}$ in the form of power series in the small parameter 
$\kappa
\delta r_{e}\ll 1$, where $\delta r_{e}$ is the amplitude of the displacement of electron in the RF field.
For the electric field  (\ref{2}) and magnetic field  (\ref{3}) this velocity is determined by the equations
\begin{eqnarray}
&\displaystyle 
\frac{\partial V_{ey}\left(z, t\right)}{\partial t}+
V_{ez}\left(z, t\right)\frac{\partial V_{ey}\left(z, t\right)}{\partial z}
\nonumber
\\ 
&\displaystyle
=-\frac{e}{m_{e}}\left(E_{0y}\left(z,t \right) 
+\frac{1}{c}V_{ez}\left(z, t\right)B_{0x}\left(z,t \right)\right),
\label{11}
\\
&\displaystyle 
\frac{\partial V_{ez}\left(z,t\right)}{\partial t}+
V_{ez}\left(z, t\right)\frac{\partial V_{ez}\left(z, t\right)}{\partial z}
\nonumber
\\ 
&\displaystyle
=\frac{e}{m_{e}c}V_{ey}\left(z, t\right)B_{0x}\left(z, t\right).
\label{12}
\end{eqnarray}
With new variables $z_{e}, t'$, determined by the relations\cite{Davidson}
\begin{eqnarray}
&\displaystyle 
z= z_{e}+\int\limits^{t'}_{0}V_{ez}\left(z_{e},t'_{1} \right) dt'_{1}, \quad t=t', 
\label{13}
\end{eqnarray}
Eqs. (\ref{11}) and (\ref{12}) becomes
\begin{eqnarray}
&\displaystyle 
\frac{\partial V_{ey}\left( z_{e},t'\right) }{\partial t'} =
-\frac{eE_{0y}}{m_{e}}e^{-\kappa \left(z_{e}+\int\limits^{t'}_{0}V_{ez}\left(z_{e},t'_{1} \right) dt'_{1}\right)}
\nonumber
\\ 
&\displaystyle
\times\left(\sin \omega_{0}t'+\frac{\kappa V_{ez}\left(z_{e},t' \right)}{\omega_{0}}\cos\omega_{0}t' \right), 
\label{14}
\end{eqnarray}
\begin{eqnarray}
&\displaystyle 
\frac{\partial V_{ez}\left( z_{e},t'\right) }{\partial t'} 
=\kappa\xi_{e0}e^{-\kappa \left(z_{e}+\int\limits^{t'}_{0}V_{ez}\left(z_{e},t'_{1} \right) dt'_{1}\right)}
\nonumber
\\ 
&\displaystyle
\times\omega_{0}V_{ey}\left(z_{e},t' \right)\cos\omega_{0}t',
\label{15}
\end{eqnarray}
where 
\begin{eqnarray}
&\displaystyle \xi_{e0}=\frac{eE_{0y}}{m_{e}\omega^{2}_{0}}
\label{16}
\end{eqnarray}
is the amplitude of the displacement of an electron along the coordinate $y$ at $z_{e}=0$. 
We find the approximate solutions to nonlinear Eqs. (\ref{14}), (\ref{15}) for $V_{ey}\left(z_{e}, t\right)$ and $V_{ez}\left(z_{e}, t\right)$  
assuming that the parameter $\kappa\xi_{e0}$ is much less than unity. In this paper, we consider the case of the high frequency $\omega_{0}$ 
RF wave for which the RF electric field force acting on 
electrons in the skin layer  prevails over the RF Lorentz force. In this case the electron cyclotron frequency formed by the magnetic field $B_{0x}$ is much 
less than $\omega_{0}$. The procedure of the solution of system (\ref{14}), (\ref{15}) for the opposite case of the low 
frequency RF wave, for which the RF Lorentz force dominates over the RF electric field force, is different and will be considered 
in the separate paper. It follows from Eq. (\ref{15}) that $V_{ez}$ is constant in zero-order approximation and without loss of the 
generality we put it to be equal to zero. In this approximation, we obtain from Eq. (\ref{14}) the equation for $V_{ey}$,
\begin{eqnarray}
&\displaystyle 
\frac{\partial V_{ey}\left(z_{e},t\right)}{\partial t}= -\frac{eE_{0y}\left(z_{e}\right)}{m_{e}}\sin \omega_{0}t,
\label{17}
\end{eqnarray}
with solution 
\begin{eqnarray}
&\displaystyle 
V_{ey}\left(z_{e},t\right)=\frac{eE_{0y}\left(z_{e}\right)}{m_{e}\omega_{0}}\cos \omega_{0}t
\nonumber
\\ 
&\displaystyle
=\frac{e}{cm_{\alpha}}A_{0y}\left(z_{e},t \right),
\label{18}
\end{eqnarray}
where $E_{0y}\left(z_{e}\right)=E_{0y}e^{-\kappa z_{e}}$ is the local value of the amplitude of the $\mathbf{E}_{0y}$ field. 
Accounting for the terms of the first order in $\kappa\xi_{e0}\ll 1$ in Eq. (\ref{15}), we find that
\begin{eqnarray}
&\displaystyle 
\frac{\partial V_{ez}\left(z_{e},t\right)}{\partial t}= \frac{e\kappa E_{0y}\left(z_{e}\right)}{m_{e}\omega_{0}}
V_{ey}\left(z_{e},t\right)\cos \omega_{0}t
\nonumber
\\ 
&\displaystyle
= \frac{e}{m_{e}}\kappa\xi_{e}E_{0y}\left(z_{e}\right)
\cos^{2} \omega_{0}t,
\label{19}
\end{eqnarray}
where 
\begin{eqnarray}
&\displaystyle \xi_{e}=\xi_{e}\left(z_{e}\right)=\xi_{e0}e^{-\kappa z_{e}}
\label{20}
\end{eqnarray}
is the amplitude of the local displacement of electron along the coordinate $y$ at $z_{e}$. Equation (\ref{19}) is similar to the equation of 
the electron motion under the action of the ponderomotive force\cite{Schmidt}, and has solution
\begin{eqnarray}
&\displaystyle
V_{ez}\left(z_{e},t \right)=\kappa\xi_{e}\frac{e }{2m_{e}}E_{0y}\left(z_{e}\right)t
\nonumber
\\ 
&\displaystyle
+\frac{1}{4}\kappa\xi_{e}\frac{e }{m_{e}\omega_{0}}E_{0y}\left(z_{e}\right)\sin2\omega_{0}t.
\label{21}
\end{eqnarray}
By employing the method of successive approximations for the solution of the nonlinear Eqs. (\ref{14}) and (\ref{15}), with $V_{ez}\left(z_{e},t \right)$
determined by Eq. (\ref{21}) as the starting approximation, we obtain for the time $t\gg \omega^{-1}_{0}$ the following equation for $V_{ey}\left(z_{e},t 
\right)$ 
\begin{eqnarray}
&\displaystyle 
\frac{\partial V_{ey}\left(z_{e},t\right)}{\partial t}= -\frac{eE_{0y}\left(z_{e}\right)}{m_{e}}e^{-\frac{1}{4}\kappa^{2}\xi^{2}_{e}\omega_{0}^{2}t^{2}}\sin 
\omega_{0}t
\label{22}
\end{eqnarray}
with solution
\begin{eqnarray}
&\displaystyle 
V_{ey}\left(z_{e},t\right)=\frac{eE_{0y}\left(z_{e}\right)}{m_{e}\omega_{0}}e^{-\frac{1}{4}\kappa^{2}\xi^{2}_{e}\omega_{0}^{2}t^{2}}\cos \omega_{0}t,
\label{23}
\end{eqnarray}
which is valid for $\kappa\xi_{e}\ll 1$. In this approximation, Eq. (\ref{15}) becomes 
\begin{eqnarray}
&\displaystyle 
\frac{\partial V_{ez}\left(z_{e},t\right)}{\partial t}
= \frac{e}{m_{e}}\kappa\xi_{e}E_{0y}\left(z_{e}\right)e^{-\frac{1}{2}\kappa^{2}\xi^{2}_{e}\omega_{0}^{2}t^{2}}
\cos^{2} \omega_{0}t,
\label{24}
\end{eqnarray}
with solution
\begin{eqnarray}
&\displaystyle
V_{ez}\left(z_{e},t \right)=\kappa\xi_{e}\frac{e }{2m_{e}}E_{0y}\left(z_{e}\right)te^{-\frac{1}{2}\kappa^{2}\xi^{2}_{e}\omega_{0}^{2}t^{2}}
\nonumber
\\ 
&\displaystyle
+\frac{1}{4}\kappa\xi_{e}\frac{e }{m_{e}\omega_{0}}E_{0y}\left(z_{e}\right)e^{-\frac{1}{2}\kappa^{2}\xi^{2}_{e}\omega_{0}^{2}t^{2}}\sin2\omega_{0}t.
\label{25}
\end{eqnarray}
Equation (\ref{25}) reveals that the acceleration maximum for $V_{ez}\left(z_{e},t \right)$ velocity attains at $t=t_{\ast}$, where 
\begin{eqnarray}
&\displaystyle 
t_{\ast}=\frac{1}{\kappa\xi_{e}\omega_{0}}.
\label{26}
\end{eqnarray} 
During time $0\leqslant t\leqslant t_{\ast}$ an electron passes the half of the skin layer depth equal to $\left(2\kappa\right)^{-1}$. 
At time $t>2t_{\ast}$, an electron enters into the inner plasma, where the EM field is exponentially small. 
Neglecting the exponentially small variation of the velocity $\mathbf{V}_{e}$ past the skin layer,
velocity $V_{ez}\left(z_{e},t \right)$ at time $t> 2t_{\ast}$ may be approximated as 
\begin{eqnarray}
&\displaystyle 
V_{ez}\left(t>t_{\ast} \right)= V_{ez}\left(t=t_{\ast} \right)=\frac{eE_{0y}}{m_{e}\omega_{0}}\exp(-2),
\label{27}
\end{eqnarray}
and $V_{ey}\left(t>t_{\ast} \right)=0$.

At time interval $0\leqslant t<t_{\ast}$, the Vlasov equation (\ref{10}) 
\begin{eqnarray}
&\displaystyle 
\frac{\partial F_{e}\left(\mathbf{v}_{e},\mathbf{r}_{e},t \right) }{\partial t}
+\mathbf{v}_{e}\frac{\partial F_{e}}{\partial\mathbf{r}_{e}}
+\frac{e}{m_{e}}\nabla \varphi \left(\mathbf{r}_{e}, t\right) \frac{\partial F_{e}}{\partial\mathbf{v}_{e}}
\nonumber
\\ 
&\displaystyle
+\kappa\xi_{e}e^{-\frac{1}{4}\kappa^{2}\xi^{2}_{e}\omega_{0}^{2}t^{2}} v_{ez}\sin\omega_{0}t\frac{\partial F_{e}}{\partial y_{e}}
\nonumber
\\ 
&\displaystyle
+\kappa \xi_{e}e^{-\frac{1}{4}\kappa^{2}\xi^{2}_{e}\omega_{0}^{2}t^{2}} v_{ey}\omega_{0}\cos\omega_{0}t\frac{\partial F_{e}}{\partial v_{ez}}=0.
\label{28}
\end{eqnarray}
governs the temporal evolution of the electron distribution function in the frame of references moving with velocity $\mathbf{V}_{e}\left(t\right)$.
At time $t>2t_{\ast}$,  Eq. (\ref{28}) may be approximated as
\begin{eqnarray}
&\displaystyle 
\frac{\partial F_{e}\left(\mathbf{v}_{e},\mathbf{r}_{e},t \right) }{\partial t}
+\mathbf{v}_{e}\frac{\partial F_{e}}{\partial\mathbf{r}_{e}}
+\frac{e}{m_{e}}\nabla \varphi \left(\mathbf{r}_{e}, t\right) \frac{\partial F_{e}}{\partial\mathbf{v}_{e}}=0.
\label{29}
\end{eqnarray}
Equations (\ref{28}), (\ref{29}) and the Vlasov equation for ions jointly with the Poisson equation (\ref{6}) for the potential 
$\varphi \left(\mathbf{r}_{e}, t\right)$ 
compose basic system of equations. It is important to note, that the spatial inhomogeneity and time dependence in the zero order in $\kappa\xi_{e}$ is 
excluded from the Maxwellian distribution (\ref{7}) in convective coordinates with velocity $V_{ey}$ determined by Eq. (\ref{17}). 
At the same time, the transition from $v_{z}$ to $v_{ez}$ introduces spatial inhomogeneity and time dependence of the first order in $\kappa\xi_{e}$ to 
$F_{e0}$. Therefore, the solution of the Vlasov equation (\ref{28}) for $F_{e0}\left(v_{ez},v_{ey},z_{e},t \right)$ may be presented 
in the form of power series in $\kappa\xi_{e}\ll 1$,
\begin{eqnarray}
&\displaystyle 
F_{e0}\left(v_{ez}, v_{ey}, z_{e}, t \right) =
F^{\left(0 \right)}_{e0}\left(v_{ez}, v_{ey}\right)
\nonumber
\\ 
&\displaystyle
+F^{\left(1 \right) }_{e0}\left(v_{ez}, v_{ey}, z_{e}, t \right),
\label{30}
\end{eqnarray}
where
\begin{eqnarray}
&\displaystyle 
F^{\left(0 \right) }_{e0}\left(v_{ez},v_{ey}\right) = \frac{n_{0e}}{2\pi v^{2}_{Te} }
\exp\left[-\frac{v^{2}_{ez}}{2v^{2}_{Te}}-\frac{v^{2}_{ey}}{2v^{2}_{Te}}\right]. 
\label{31}
\end{eqnarray}
With expansion (\ref{30}) the spatial inhomogeneity and time dependence of $F_{e0}\left(v_{ez}, v_{ey}, z_{e}, t \right)$ 
in the convective coordinates is determined by $F_{e0}^{\left(1 \right)}$, which is the solution of Eq. (\ref{28}) with 
$\varphi \left(\mathbf{v}_{e},\mathbf{r}_{e}, t\right) =0$,
\begin{eqnarray}
&\displaystyle 
\frac{\partial F_{e0}^{\left(1 \right)} }{\partial t}
+v_{ez}\frac{\partial F_{e0}^{\left(1 \right) } }{\partial z_{e}}
\nonumber
\\ 
&\displaystyle
=-\kappa\xi_{e}\omega_{0}e^{-\frac{1}{4}\kappa^{2}\xi^{2}_{e}\omega_{0}^{2}t^{2}}\frac{v_{ey}}{2}\left(e^{-i\omega_{0}t} + e^{i\omega_{0}t}\right)
\frac{\partial F_{e0}^{\left(0 \right)}}{\partial v_{ez}}.
\label{32}
\end{eqnarray}
With new characteristic variable $z'_{e}=z_{e}-v_{ez}t$, the derivative over $z_{e}$ is excluded from Eq. (\ref{32}) and the solution to Eq. (\ref{32}) 
becomes
\begin{eqnarray}
&\displaystyle 
F_{e0}^{\left(1 \right)}\left(\mathbf{v}_{e},z',t \right) = -\kappa\xi_{e}\omega_{0}e^{-\frac{1}{4}\kappa^{2}\xi^{2}_{e}\omega_{0}^{2}t^{2}}\frac{v_{ey}}{2}
\frac{\partial  F_{e0}^{\left(0 \right)} }{\partial v_{ez}}
\nonumber
\\ 
&\displaystyle
\times \left(\frac{e^{i\omega_{0} t}}{i\omega_{0} -\kappa v_{ez}}-\frac{e^{-i\omega_{0} t}}
{i\omega_{0} +\kappa v_{ez}} \right)+\Psi\left(v_{ez},v_{ey} \right).
\label{33}
\end{eqnarray}
The function $\Psi\left(v_{ez},v_{ey} \right)$ is determined by employing simple boundary conditions\cite{Shaing} determined for different values of  
coordinate $z_{e}$. The first condition is applied at $z_{e}=\infty$ for the electrons moving from  $z_{e}=\infty$ toward plasma boundary  $z_{e}=0$, i. e. 
for electrons with velocity $v_{ez}<0$. Because the electric field $E_{0y}\left(z_{e}\right)$ vanishes at $z_{e}=\infty$, the boundary condition $F_{e0}
^{\left(1\right)}\left(\mathbf{v}_{e},z'\rightarrow +\infty, t \right) =0$ determines $\Psi\left(v_{ez}<0,v_{ey} \right)=0$ and  
\begin{eqnarray}
&\displaystyle 
F_{e0}^{\left(1 \right)}\left(v_{ey}, v_{ez}<0, z', t \right) =
-\kappa\xi_{e}\omega_{0}\frac{v_{ey}}{2}
\frac{\partial F_{e0}^{\left(0 \right)}}{\partial v_{ez}}
\nonumber
\\ 
&\displaystyle
\times e^{-\frac{1}{4}\kappa^{2}\xi^{2}_{e}\omega_{0}^{2}t^{2}}\left(\frac{e^{i\omega_{0} t}}{i\omega_{0} -\kappa v_{ez}}
-\frac{e^{-i\omega_{0} t}}{i\omega_{0} +\kappa v_{ez}}
\right). 
\label{34}
\end{eqnarray}
The second boundary condition is the condition of the specular reflection of electrons  at the plasma boundary $z=0$,
\begin{eqnarray}
&\displaystyle 
F_{e0}^{\left(1 \right)}\left(v_{ey}, v_{ez}<0, z=0, t\right)
\nonumber
\\ 
&\displaystyle
= F_{e0}^{\left(1 \right)}\left(v_{ey}, v_{ez}>0, z=0, t\right).
\label{35}
\end{eqnarray}
This condition determines the solution for electron distribution function  $F_{e0}^{\left(1 \right)}\left(v_{ey}, v_{ez}>0,z',t \right)$ in a form
\begin{eqnarray}
&\displaystyle 
F_{e0}^{\left(1 \right)}\left(v_{ey}, v_{ez}>0,z',t \right) =
\nonumber
\\ 
&\displaystyle
-\kappa\omega_{0}e^{-\frac{1}{4}\kappa^{2}\xi^{2}_{e}\omega_{0}^{2}t^{2}}\frac{v_{ey}}{2}
\frac{\partial F_{e0}^{\left(0 \right)}}{\partial v_{ez}}
\nonumber
\\ 
&\displaystyle
\times
\left[\xi_{e}\left(\frac{e^{i\omega_{0} t}}{i\omega_{0} -\kappa v_{ez}}
-\frac{e^{-i\omega_{0} t }}{i\omega_{0} +\kappa v_{ez}}
\right)
e^{-\kappa v_{ez}t }\right.
\nonumber
\\ 
&\displaystyle
\left.
+\frac{4\kappa \xi_{e0} v_{ez}}{\omega_{0}^{2}+\kappa^{2} v_{ez}^{2}}\cos \omega_{0} t\right].  
\label{36}
\end{eqnarray}
With the equilibrium distribution function $F_{e0}$, determined by Eq. (\ref{31}), 
the Vlasov equation (\ref{28}) for the perturbation $f_{e}$ of the electron distribution function becomes
\begin{eqnarray}
&\displaystyle 
\frac{\partial f_{e}\left(\mathbf{v}_{e},\mathbf{r}_{e},t \right) }{\partial t}
+\mathbf{v}_{e}\frac{\partial f_{e}}{\partial\mathbf{r}_{e}}
\nonumber
\\ 
&\displaystyle
+\kappa\xi_{e}e^{-\frac{1}{4}\kappa^{2}\xi^{2}_{e}\omega_{0}^{2}t^{2}}v_{ez}\sin\omega_{0}t\frac{\partial f_{e}}{\partial y_{e}}
\nonumber
\\ 
&\displaystyle
+\kappa \xi_{e}e^{-\frac{1}{4}\kappa^{2}\xi^{2}_{e}\omega_{0}^{2}t^{2}} v_{ey}\omega_{0}\cos\omega_{0}t\frac{\partial f_{e}}
{\partial v_{ez}}
\nonumber
\\ 
&\displaystyle
+\frac{e}{m_{e}}\nabla \varphi \left(\mathbf{r}_{e}, t
\right)\frac{\partial }{\partial\mathbf{v}_{e}}\left( F^{\left(0 \right)}_{e0}\left(v_{ez},v_{ey}\right)\right.
\nonumber
\\ 
&\displaystyle
\left.+F^{\left(1 \right) }_{e0}\left(v_{ez}, v_{ey}, z_{e}, t \right)+f_{e}\left(\mathbf{v}_{e},\mathbf{r}_{e},t \right)\right) =0,
\label{37}
\end{eqnarray}
which contains the electrostatic potential $\varphi \left(\mathbf{r}_{e}, t\right)$ of the self-consistent 
respond of a plasma on the RF wave. The solution to Eq. (\ref{37}) may be found in the form of power series in $\kappa\xi_{e}\ll 1$.

\section{Electron convecting-ocsillating mode}\label{sec3}

In the zero order in $\kappa\xi_{e}$, the equilibrium distribution functions $F_{e0,i0}$ in the convective 
coordinates are determined by the spatially homogeneous functions $F^{(0)}_{e0,i0}\left(\mathbf{v}_{e,i}\right)$. 
In this approximation, Eq. (\ref{37})  for $f_{e}\left(\mathbf{v}_{e}, \mathbf{r}_{e}, t \right)$ 
and similar equation for $f_{i}\left(\mathbf{v}_{i},\mathbf{r}_{i},t \right)$ 
do not contain the RF electric field in their convective-oscillating frames. Therefore the equations for $f_{i}$ and $f_{e}$ 
will be the same as for the plasma without RF field,
\begin{eqnarray}
&\displaystyle 
\frac{\partial f_{i}}{\partial
t}+\mathbf{v}_{i}\frac{\partial f_{i}}
{\partial\mathbf{r}_{i}}-\frac{e_{i}}{m_{i}}\nabla 
\varphi_{i}\left(\mathbf{r}_{i},t\right)\frac{\partial
F_{i0}}{\partial\mathbf{v}_{i}}=0,
\label{38}
\\
&\displaystyle \frac{\partial f_{e}}{\partial
t}+\mathbf{v}_{e}\frac{\partial f_{e}}
{\partial\mathbf{r}_{e}}-\frac{e}{m_{e}}\nabla 
\varphi_{e}\left(\mathbf{r}_{e},t\right)\frac{\partial
F_{e0}}{\partial\mathbf{v}_{e}}=0.
\label{39}
\end{eqnarray}
The solution of the linearised equations for $f_{i}$ Fourier transformed over $\mathbf{r}_{i}$ is
\begin{eqnarray}
&\displaystyle 
f_{i}\left(\mathbf{v}_{i}, \mathbf{k}_{i}, t\right)=i\frac{e_{i}}{m_{i}}\mathbf{k}_{i}\frac{\partial F_{i0}}{\partial \mathbf{v}_{i}}\int 
\limits^{t}_{0}dt_{1}\varphi_{i}\left(\mathbf{k}_{i}, t_{1}\right)
\nonumber
\\ 
&\displaystyle
\times
e^{-i\mathbf{k}_{i}\mathbf{v}_{i}\left(t-t_{1}\right)},
\label{40}
\end{eqnarray}
where $\varphi_{i}\left(\mathbf{k}_{i}, t_{1}\right)$ is the Fourier transform of the potential 
$\varphi_{i}\left(\mathbf{r}_{i}, t_{1}\right)$ over $\mathbf{r}_{i}$,
\begin{eqnarray}
&\displaystyle 
\varphi_{i}\left(\mathbf{k}_{i}, t_{1}\right)=\frac{1}{\left(2\pi\right)^{3}}\int 
d\mathbf{r}_{i}\varphi_{i}\left(\mathbf{r}_{i}, t_{1}\right)e^{-i\mathbf{k}_{i}\mathbf{r}_{i}}.
\label{41}
\end{eqnarray}
The ion density perturbation $n_{i}\left(\mathbf{k}_{i}, t\right)$ Fourier transformed over $\mathbf{r}_{i}$ with the conjugate wave vector
$\mathbf{k}_{i}$  is
\begin{eqnarray}
&\displaystyle 
n_{i}\left(\mathbf{k}_{i}, t\right)=\int f_{i}\left(\mathbf{v}_{i}, \mathbf{k}_{i}, t\right)d\mathbf{v}_{i}
=i\frac{e_{i}}{m_{i}}\mathbf{k}_{i}
\nonumber
\\ 
&\displaystyle
\times\int d\mathbf{v}_{i} \frac{\partial F_{i0}}{\partial \mathbf{v}_{i}}\int 
\limits^{t}_{0}dt_{1}\varphi_{i}\left(\mathbf{k}_{i}, t_{1}\right)e^{-i\mathbf{k}_{i}\mathbf{v}_{i}\left(t-t_{1}\right)}.
\label{42}
\end{eqnarray}
The Fourier transform $n_{e}\left(\mathbf{k}_{e}, t\right)$ of the electron density 
perturbation performed in the electron frame is given 
by equation
\begin{eqnarray}
&\displaystyle 
n_{e}\left(\mathbf{k}_{e}, t\right)=\int f_{e}\left(\mathbf{v}_{e}, \mathbf{k}_{e}, t\right)d\mathbf{v}_{e}
=i\frac{e}{m_{e}}\mathbf{k}_{e}
\nonumber
\\ 
&\displaystyle
\times\int d\mathbf{v}_{e} \frac{\partial F_{e0}}{\partial \mathbf{v}_{e}}\int 
\limits^{t}_{0}dt_{1}\varphi_{e}\left(\mathbf{k}_{e}, t_{1}\right)e^{-i\mathbf{k}_{e}\mathbf{v}_{e}\left(t-t_{1}\right)}.
\label{43}
\end{eqnarray}
which is the same as Eq. (\ref{42}) for $n_{i}\left(\mathbf{k}_{i}, t\right)$ with changing ion on electron subscripts.

The perturbations  of the ion, (\ref{42}), and electron, (\ref{43}), densities are used in the Poisson equation (\ref{6})
which may be the equation for $\varphi_{i}\left(\mathbf{k}_{i}, t_{1}\right)$ by the Fourier transform of Eq. (\ref{6}) over $\mathbf{r}_{i}$,
\begin{eqnarray}
& \displaystyle 
k^{2}_{i}\varphi_{i}\left(\mathbf{k}_{i},t\right)=4\pi e\left(n_{i}\left(\mathbf{k}_{i}, t\right)\right.
\nonumber
\\ 
&\displaystyle
\left.-\int d\mathbf{r}_{i}n_{e}\left(\mathbf{r}_{e}, t\right)e^{-i\mathbf{k}_{i}\mathbf{r}_{i}}\right),
\label{44}
\end{eqnarray}
or as the equation for $\varphi_{e}\left(\mathbf{k}_{e},t\right)$ by the Fourier transform of Eq. (\ref{6}) over $\mathbf{r}_{e}$. 
For the deriving the Poisson equation for $\varphi_{i}\left(\mathbf{k}_{i} , t\right)$ the Fourier transforms 
$n^{(i)}_{e}\left(\mathbf{k}_{i}, t\right)$ and 
$\varphi^{(i)}_{e}\left(\mathbf{k}_{i}, t\right)$ of $n_{e}\left(\mathbf{r}_{e}, t\right)$ 
and $\varphi_{e}\left(\mathbf{r}_{e}, t\right)$ 
over $\mathbf{r}_{i}$ should be determined. Using Eq. (\ref{9}), which determines the relations among the coordinates in the 
laboratory, ion and electron frames, we find that the electron density perturbation $n_{e}\left(\mathbf{r}_{e}, t\right)$ 
Fourier transformed over $\mathbf{r}_{i}$ is
\begin{eqnarray}
&\displaystyle
n^{(i)}_{e}\left(\mathbf{k}_{i}, t\right)=\int d\mathbf{r}_{i}n_{e}\left(\mathbf{r}_{e}, t\right)e^{-i\mathbf{k}_{i}\mathbf{r}_{i}}=\int d
\mathbf{r}_{e}n_{e}\left(\mathbf{r}_{e}, t\right)
\nonumber
\\ 
&\displaystyle
\times 
\exp\left(-i\mathbf{k}_{i}\mathbf{r}_{e}-ik_{iy}\int\limits^{t}_{0} dt_{1}\left(V_{ey}\left(t_{1}\right)-V_{iy}\left(t_{1}\right)\right)\right.
\nonumber
\\ 
&\displaystyle
\left.-ik_{iz}\int\limits^{t}_{0} dt_{1}\left(V_{ez}\left(t_{1}\right)-V_{iz}\left(t_{1}\right)\right)\right),
\label{45}
\end{eqnarray}
where velocities $V_{ey}\left(t_{1}\right)$ and $V_{ez}\left(t_{1}\right)$ are determined by Eqs. (\ref{23}) and (\ref{25}). The velocities $V_{iy}
\left(t_{1}\right)$ and $V_{iz}\left(t_{1}\right)$, which are determined by the same Eqs. (\ref{23}) and (\ref{25}) with subscript $i$ 
instead of $e$, are in $m_{i}/m_{e}$  times less than $V_{ey}$ and $V_{ez}$ and are neglected in what follows. 
With velocities $V_{ey}\left(t\right)$ and $V_{ez}\left(t\right)$ determined by Eqs. (\ref{19}) and (\ref{21}) relation (\ref{43})
for time $t< t_{\ast}$ becomes
\begin{eqnarray}
&\displaystyle 
n^{(i)}_{e}\left(\mathbf{k}_{i}, t\right)=n^{(e)}_{e}\left(\mathbf{k}_{i}, t\right)\exp\left(-ik_{iy}\xi_{e}\sin\omega_{0}t
\right. 
\nonumber
\\ 
&\displaystyle
\left. 
-i\frac{1}{2}k_{iz}a_{e}t^{2} +ik_{iz}\eta_{e}\cos 2\omega_{0}t\right)
\label{46}
\end{eqnarray}
where $\xi_{e}$ is determined by Eq. (\ref{20}), and 
\begin{eqnarray}
&\displaystyle 
\eta_{e}=\eta_{e}\left(z_{e}\right)=\frac{1}{8}\kappa\xi^{2}_{e}\left(z_{e}\right)
\label{47}
\end{eqnarray}
is the amplitude of the electron displacement in the RF electric field along coordinate $z_{e}$, and 
\begin{eqnarray}
&\displaystyle 
a_{e}= \frac{1}{2}\frac{e^{2}\kappa E^{2}_{0y}}{m^{2}_{e}\omega^{2}_{0}}
\label{48}
\end{eqnarray}
is the electron acceleration under the action of the ponderomotive force.

The relation between the Fourier transform $\varphi_{e}\left( \mathbf{k}_{e}, t\right)$ of 
the potential $\varphi_{e}\left( \mathbf{r}_{e}, t\right)$ over 
$\mathbf{r}_{e}$, involved in Eq. (\ref{41}) 
for $n_{e}\left(\mathbf{r}_{e}, t\right)$, and the Fourier transform $\varphi_{i}\left( 
\mathbf{k}_{i}, t\right)$ of the potential $\varphi_{e}\left( 
\mathbf{r}_{e}, t\right)$ over $\mathbf{r}_{i}$ when it is 
used in $n^{(i)}_{e}\left(\mathbf{k}_{i}, t\right)$, is derived similar and is determined by 
the relation
\begin{eqnarray}
&\displaystyle 
\varphi^{(e)}_{e}\left(\mathbf{k}_{e},t_{1}\right)=\exp\left(ik_{iy}\xi_{e}
\sin \omega_{0}t_{1}+i\frac{1}{2}k_{iz}a_{e}t_{1}^{2}\right.
\nonumber
\\ 
&\displaystyle
\left.-ik_{iz}\eta_{e}\cos 
2\omega_{0}t_{1}\right)\varphi_{i}\left(\mathbf{k}_{i}, t_{1}\right),
\label{49}
\end{eqnarray}
which follows from the identity $\varphi_{e}\left(\mathbf{r}_{e},t_{1}\right)=\varphi_{i}
\left(\mathbf{r}_{i},t_{1}\right)$.

With Eq. (\ref{40}) for $n_{i}\left(\mathbf{k}_{i}, t\right)$, and with Eq. (\ref{46}) for $n^{(i)}_{e}\left(\mathbf{k}_{i}, t\right)$ in 
which potential is determined by (\ref{49}), the Poisson equation (\ref{44}) gives the following equation 
\begin{eqnarray}
&\displaystyle
k^{2}_{i}\varphi_{i}\left(\mathbf{k}_{i}, t\right)=\frac{4\pi i e^{2}_{i}}{m_{i}}
\int d\mathbf{v}_{i}\mathbf{k}_{i}\frac{\partial F_{i0}}{\partial \mathbf{v}_{i}}
\nonumber
\\ 
&\displaystyle
\times\int \limits^{t}_{0}dt_{1}\varphi_{i}\left(\mathbf{k}_{i}, t_{1}\right)e^{-i\mathbf{k}_{i}\mathbf{v}_{i}\left(t-t_{1}\right)}
\nonumber
\\ 
&\displaystyle
+\frac{4\pi i e^{2}_{i}}{m_{e}}
\int d\mathbf{v}_{e}\mathbf{k}_{i}\frac{\partial F_{e0}}{\partial \mathbf{v}_{e}}
\int \limits^{t}_{0}dt_{1}\varphi_{i}\left(\mathbf{k}_{i}, t_{1}\right)
\nonumber
\\ 
&\displaystyle
\exp\left(-i\mathbf{k}_{i}\mathbf{v}_{e}\left(t-t_{1}\right)-ik_{iy}\xi_{e}
\left(\sin \omega_{0}t-\sin \omega_{0}t_{1}\right)\right.
\nonumber
\\ 
&\displaystyle
\left.-i\frac{1}{2}k_{iz}a_{e}\left(t^{2}-t_{1}^{2}\right)+ik_{iz}\eta_{e}\left(\cos 
2\omega_{0}t-\cos 2\omega_{0}t_{1}\right)\right),
\label{50}
\end{eqnarray}
which determines  the evolution of the  electrostatic potential $\varphi_{i}\left(\mathbf{k}
_{i}, \omega\right)$ in the skin layer of an inductively coupled plasma for time $t< t_{\ast}$.
For the Maxwellian distribution $F_{0\alpha}
\left(\mathbf{v}_{\alpha}\right)$, determined by Eq. (\ref{31}) for electrons and ions,
this equation becomes
\begin{eqnarray}
&\displaystyle
\varphi_{i}\left(\mathbf{k}_{i}, t\right)+\omega^{2}_{pi}
\int \limits^{t}_{0}dt_{1}\varphi_{i}\left(\mathbf{k}_{i}, t_{1}\right)\left(t-t_{1}\right)
e^{-\frac{1}{2}k^{2}_{i}v^{2}_{Ti}\left(t-t_{1}\right)^{2}}
\nonumber
\\ 
&\displaystyle
+\omega^{2}_{pe}\int \limits^{t}_{0}dt_{1}\varphi_{i}\left(\mathbf{k}_{i}, t_{1}\right)\left(t-t_{1}\right)
e^{-\frac{1}{2}k^{2}_{i}v^{2}_{Te}\left(t-t_{1}\right)^{2}}
\nonumber
\\ 
&\displaystyle
\times\exp\left(-i\frac{1}{2}k_{iz}a_{e}\left(t^{2}-t_{1}^{2}\right)-ik_{iy}\xi_{e}\left(\sin \omega_{0}t-\sin \omega_{0}t_{1}\right)\right.
\nonumber
\\ 
&\displaystyle
\left.+ik_{iz}\eta_{e}\left(\cos 2\omega_{0}t-\cos 2\omega_{0}t_{1}\right)\right)=0.
\label{51}
\end{eqnarray}
It follows from Eq. (\ref{51}) for $k_{iz}\sim k_{iy}$, that because 
\begin{eqnarray}
&\displaystyle
\frac{k_{iy}\xi_{e}}{k_{iz}\eta_{e}}\sim \frac{8}{\kappa\xi_{e}}> 1,
\label{52}
\end{eqnarray}
and 
\begin{eqnarray}
&\displaystyle
\frac{a_{e}t^{2}}{2\xi_{e}}\sim \kappa\xi_{e}\omega^{2}_{0}t^{2}>1,
\label{53}
\end{eqnarray} 
when 
\begin{eqnarray}
&\displaystyle
1>\frac{t}{t_{\ast}}>\frac{1}{\omega_{0}t},
\label{54}
\end{eqnarray} 
the uniformly accelerating motion of electrons, which stems from the ponderomotive force, dominates over their oscillating motion. Therefore 
the possible instability of the skin layer under condition $\kappa\xi_{e}< 1$ at time interval (\ref{54}) is the current driven instability 
with accelerated electron current velocity, instead of the supposed instability of the parametric type. At the larger time, $t > 2t_{\ast}$, the 
instabilities may be developed outside the skin layer, where Eq. (\ref{44}) for $\varphi_{i}\left(\mathbf{k}_{i}, t\right)$ has a form
\begin{eqnarray}
&\displaystyle
\varphi_{i}\left(\mathbf{k}_{i}, t\right)+\omega^{2}_{pi}
\int \limits^{t}_{0}dt_{1}\varphi_{i}\left(\mathbf{k}_{i}, t_{1}\right)\left(t-t_{1}\right)
e^{-\frac{1}{2}k^{2}_{i}v^{2}_{Ti}\left(t-t_{1}\right)^{2}}
\nonumber
\\ 
&\displaystyle
+\omega^{2}_{pe}\int \limits^{t}_{0}dt_{1}\varphi_{i}\left(\mathbf{k}_{i}, t_{1}\right)\left(t-t_{1}\right)
\nonumber
\\ 
&\displaystyle
\times e^{-\frac{1}{2}k^{2}_{i}v^{2}_{Te}\left(t-t_{1}\right)^{2}-ik_{iz}V_{ez}\left(t-t_{1}\right)}=0,
\label{55}
\end{eqnarray}
where $V_{ez}$ is determined by the estimate (\ref{27}).

\section{The instabilities driven by the electrons accelerated in the skin layer}\label{sec4} 

In this section, we derive the solution to Eq. (\ref{51}) for potential $\varphi_{i}\left(\mathbf{k}_{i}, t\right)$ in the time interval (\ref{54}). We will find the solution to Eq. (\ref{51}) in the WKB-like form
\begin{eqnarray}
&\displaystyle 
\varphi\left(\mathbf{k}_{i}, t_{1} \right) =\varphi_{i}\left(\mathbf{k}_{i}\right)
e^{-i \int\limits^{t_{1}}_{0}\omega\left(\mathbf{k}_{i}, t_{2} \right)dt_{2}}, 
\label{56}
\end{eqnarray} 
where $\varphi\left(\mathbf{k}_{i}\right)=\int \limits^{\infty}_{-\infty}e^{i\mathbf{k}_{i}
\mathbf{r}_{i}}\varphi\left(\mathbf{r}_{i}, 0\right)d\mathbf{r}_{i}$ is the Fourier 
transform of the initial perturbation of $\varphi\left(\mathbf{r}_{i}, t_{1} \right)$ at $ t_{1}=0$. Then, Eq. (\ref{51}) becomes 
\begin{eqnarray}
&\displaystyle 
\varphi\left(\mathbf{k}_{i}\right)\left[1+\omega^{2}_{pi}\int \limits^{t}_{0}dt_{1}\left(t-t_{1}\right)\right.
\nonumber
\\ 
&\displaystyle
\left.\times e^{-i\int\limits^{t}_{t_{1}}\omega\left(\mathbf{k}_{i}, t_{2}\right)dt_{2}-\frac{1}{2}k^{2}_{i}v^{2}_{Ti}\left(t-t_{1}\right)^{2}}\right.
\nonumber
\\ 
&\displaystyle
\left.+\omega^{2}_{pe}\int \limits^{t}_{0}dt_{1}\left(t-t_{1}\right)
\right.
\nonumber
\\ 
&\displaystyle
\left.
\times e^{-i\int\limits^{t}_{t_{1}}\left(\omega\left(\mathbf{k}_{i}, t_{2}\right)
+ik_{iz}a_{e}t_{2}\right)dt_{2}-\frac{1}{2}k^{2}_{i}v^{2}_{Te}\left(t-t_{1}\right)^{2}}\right]=0.
\label{57}
\end{eqnarray} 
It is well known\cite{Buneman,Akhiezer1} that in plasmas with steady uniform flow of electron relative to ions the maximum growth rate have the hydrodynamic 
current driven instabilities, which develop when the electron current velocity exceeds the electron thermal velocity. Here we consider the possibility of the 
development of these instabilities in the narrow skin layer by the electron current with accelerated current velocity. For this goal we derive the 
asymptotics  of the ion and electron terms of Eq. (\ref{57}) in the hydrodynamic limit corresponding to the weak electron and ion Landau 
damping, for which $|\omega\left(\mathbf{k}_{i}, t_{1}\right)|\gg k_{i}v_{Ti}$ and 
\begin{eqnarray}
&\displaystyle
|\omega\left(\mathbf{k}_{i}, t_{1}\right)+k_{iz}a_{e}t_{1}|\gg k_{i}v_{Te}.
\label{58}
\end{eqnarray}
By integration by parts employing the presentation
\begin{eqnarray}
&\displaystyle
e^{-i\int\limits^{t}_{t_{1}}\omega\left(\mathbf{k}_{i}, t_{1}\right)dt_{1}}=\frac{i}{\omega\left(\mathbf{k}_{i}, t_{1}\right)}\frac{d}{dt_{1}}
\left(e^{-i\int\limits^{t}_{t_{1}}\omega\left(\mathbf{k}_{i}, t_{2}\right)dt_{2}}\right)
\label{59}
\end{eqnarray} 
in the ion term and the similar presentation for $e^{i\int\limits^{t}_{t_{1}}\left(\omega\left(\mathbf{k}_{i}, t_{2}\right)
+ik_{iz}a_{e}t_{2}\right)dt_{2}}$ in the electron term, we derive the equation
\begin{eqnarray}
&\displaystyle
\left[1-\frac{\omega^{2}_{pi}}{\omega^{2}\left(\mathbf{k}_{i}, t\right)}-\frac{\omega^{2}_{pe}}{\left(\omega\left(\mathbf{k}_{i}, t\right)+k_{iz}a_{e}t
\right)^{2}}\right]
\nonumber
\\ 
&\displaystyle
=Q\left(\mathbf{k}_{i}, t, t=0\right),
\label{60}
\end{eqnarray} 
where $Q\left(\mathbf{k}_{i}, t, t=0\right)= e^{i\int\limits^{t}_{0}\omega\left(\mathbf{k}_{i}, t_{1}\right)dt_{1}}q\left(\mathbf{k}_{i}, t=0\right)$
originates from the limit $t=0$ of the integration by parts of Eq. (\ref{57}). For the potential exponentially growing with time, for which 
$\text{Im}\,\omega\left(\mathbf{k}_{i}, t_{1}\right)>0$, the function $Q\left(\mathbf{k}_{i}, t, t=0\right)$ is exponentially small and may be neglected. 
Then, the left hand side of Eq. (\ref{60}) forms the equation for the time dependent frequency $\omega\left(\mathbf{k}_{i}, t\right)$. This equation is 
similar to the well known dispersion equation for the hydrodynamic electrostatic instabilities\cite{Buneman,Akhiezer1} for a plasma in which electrons are 
moving relative to ions with uniform steady velocity. The solution to Eq. (\ref{60}) with maximum value $\gamma_{max}$ of the growth rate $\gamma
\left(\mathbf{k}_{i}, t\right)= \text{Im}\,\omega\left(\mathbf{k}_{i}, t\right)$ occurs for the Buneman instability\cite{Buneman}. This instability in 
plasmas with accelerated electrons develops\cite{Akhiezer1} for perturbations with frequency 
\begin{eqnarray}
&\displaystyle
|\omega\left(\mathbf{k}_{i}, t\right)| \ll |k_{iz}a_{e}t|
\label{61}
\end{eqnarray}
under the resonance condition\cite{Akhiezer1}
\begin{eqnarray}
&\displaystyle
\omega_{pe}\approx|k_{iz}a_{e}t|.
\label{62}
\end{eqnarray}
The growth gate of the Buneman instability in this case is time dependent and is equal to 
\begin{eqnarray}
&\displaystyle
\gamma\left(\mathbf{k}_{i}, t\right)=\gamma_{max}=\frac{\sqrt{3}}{2^{4/3}}\omega_{pe}\left(\frac{m_{e}}{m_{i}}\right)^{1/3}
\nonumber
\\ 
&\displaystyle
\approx\frac{\sqrt{3}}{2^{4/3}}|k_{z}|a_{e}t
\left(\frac{m_{e}}{m_{i}}\right)^{1/3}.
\label{63}
\end{eqnarray}
It follows from Eq. (\ref{58}) that this instability develops when $a_{e}t> v_{Te}$, i. e. when
\begin{eqnarray}
&\displaystyle
E_{0y}>2\omega_{0}v_{Te}\frac{m_{e}}{e}\frac{t_{\ast}}{t}.
\label{64}
\end{eqnarray}
Equation (\ref{63}) is valid for the finite time $\bigtriangleup t$, for which the resonance condition (\ref{63}) not have time 
to be destroyed because of the acceleration of electrons, i. e. when 
\begin{eqnarray}
&\displaystyle
|k_{iz}a_{e}|\bigtriangleup t\lesssim \gamma_{max}.
\label{65}
\end{eqnarray} 

For the illustration purposes we present the numerical estimates for Eqs. (\ref{61}) - (\ref{65}) for the argon plasma 
$\left(\left(m_{Ar}/m_{e}\right)^{1/3}=42\right)$ with density $n_{0e}=10^{11} cm^{-3}$, 
electron temperature $T_{e}=3\,eV$, electric field $E_{0y}=15\,V/cm$, $\omega_{0}=10^{-2}\omega_{pe}=1.7\
\cdot 10^{8}\,sec^{-1}$, and $\kappa^{-1}=L_{s}=c/\omega_{pe}=1.8\,cm$. For these plasma and RF field parameters 
$t_{\ast}=1.2\cdot 10^{-8}\,sec$, $\gamma_{max}=2.9\cdot 10^{8}sec^{-1}$ and $\gamma_{max}t_{\ast}\approx 3.5> 1$. Condition 
(\ref{63}) for these parameters determines the $k_{iz}$ value,
\begin{eqnarray}
&\displaystyle
k_{iz}=\frac{\omega_{pe}}{a_{e}t_{\ast}}\frac{t_{\ast}}{t},
\label{66}
\end{eqnarray}
which gives for the employed parameters the estimate $k_{iz}\approx 210\frac{t_{\ast}}{t}cm^{-1}$. For $k_{iz}=240\,cm^{-1}$ we found that the resonance 
condition  (\ref{62}) is violated within a time $\bigtriangleup t\sim 
1.8\cdot 10^{-10}sec\ll t_{\ast}$ for which $\gamma_{max} \bigtriangleup t\approx 4,2\cdot 10^{-2}\ll 1$. Thus, the developed simplified theory of the Buneman instability valid for the time $t<t_{\ast}$ appeared to be sufficient for the prediction that the
non-modal Buneman instability of the skin layer driven by the accelerated electrons has not sufficient time for the linear growth and transition to 
the nonlinear stage for these plasma and RF field parameters. 

Now, we consider the non-resonant oscillations for which 
\begin{eqnarray}
&\displaystyle
|\omega_{pe}|> |k_{iz}a_{e}t|.
\label{67}
\end{eqnarray}
At time $t$, this condition occurs for  $k_{iz}$ values less than ones determined by Eq. (\ref{66}).
The growth rate $\gamma\lesssim \omega_{pi}$ for these perturbations is equal to\cite{Akhiezer1}
\begin{eqnarray}
&\displaystyle
\gamma\left(\mathbf{k}_{i}, t\right)=\omega_{pi}\left(\frac{\omega^{2}_{pe}}{k^{2}_{iz}a^{2}_{e}t^{2}}-1\right)^{-1/2}< \omega_{pi}.
\label{68}
\end{eqnarray}
For the above employed plasma parameters, the growth rate (\ref{68}) is approximately equal to
\begin{eqnarray}
&\displaystyle
\gamma\left(\mathbf{k}_{i}, t\right)\approx k_{iz}a_{e}t_{\ast}\left(\frac{m_{e}}{m_{i}}\right)^{1/2}\frac{t}{t_{\ast}}
\nonumber
\\ 
&\displaystyle
\approx k_{iz}\frac{eE_{0y}}{2m_{e}\omega_{0}}\left(\frac{m_{e}}{m_{i}}\right)^{1/2}\frac{t}{t_{\ast}}.
\label{69}
\end{eqnarray}
For $k_{iz}=2\cdot 10^{2}\,cm^{-1}$ and $E_{0y}=15 V/cm$, $\gamma\left(\mathbf{k}_{i}, t_{\ast}\right)= 5.5\cdot 10^{7}\,sec^{-1}$ and $\gamma
\left(\mathbf{k}_{i}, t_{\ast}\right)t_{\ast}\approx 0.7$. Eq. (\ref{68}) reveals that this instability exists during finite time at which condition 
(\ref{67}) holds; it transforms to the Buneman instability at time for which resonance condition (\ref{62}) occurs.  The transition time 
to the Buneman instability $t_{tr}=\omega_{pe}/k_{iz}a_{e}$ for the employed parameters is equal approximately to $1.3\cdot 10^{-8} \,sec$. Thus, after the 
time less than inverse growth rate (\ref{68}), this non - resonant instability  transforms to the Buneman instability, which in turns also has not a time for 
the development due to the violation of the resonance condition (\ref{66}). Thus, the Buneman and the non-resonant hydrodynamic instabilities can't develop 
in the normal skin layer. 

The severe restrictions, imposed by Eqs. (\ref{61}), (\ref{62}), (\ref{65}) on the development of the hydrodynamic instabilities in the skin layer, are 
absent for the bulk of plasma past the skin layer. In this region, the RF field is exponentially small and electrons move relative to ions with steady 
uniform velocity $V_{ez}$ determined by Eq. (\ref{27}), which may be much less than the electron thermal velocity $v_{Te}$. In such a plasma with cold ions $
\left(T_{e}>T_{i}\right)$, the ion-acoustic current-driven  instability\cite{Akhiezer1} develops due to the inverse electron Landau damping of the ion -
acoustic waves, when the electron velocity $V_{ez}$ is above the ion-acoustic velocity $v_{s}=\left(T_{e}/m_{i}\right)^{1/2}$. 
The dispersion equation for the frequency $\omega\left(\mathbf{k}_{i}\right)$ in 
this case is derived  easily from Eq. (\ref{55}) and is equal to
\begin{eqnarray}
&\displaystyle
1+\frac{1}{k^{2}_{i}\lambda^{2}_{De}}\left(1+i\sqrt{\frac{\pi}{2}}\frac{\left(\omega-k_{iz}V_{ez}\right)}{k_{i}v_{Te}}\right)
-\frac{\omega^{2}_{pi}}{\omega^{2}}
\nonumber
\\ 
&\displaystyle
+i\sqrt{\frac{\pi}{2}}\frac{\omega^{2}_{pi}\omega}{k^{3}v^{3}_{Ti}}\exp\left(-\frac{\omega^{2}}{2k^{2}v^{2}_{Ti}}\right)=0,
\label{70}
\end{eqnarray}
where $\lambda _{De}$ is the electron Debye radius. The frequency $\omega\left(\mathbf{k}_{i}\right)$ and the growth rate $\gamma\left(\mathbf{k}_{i}\right)$ 
of the ion-acoustic current-driven instability are equal to 
\begin{eqnarray}
&\displaystyle
\omega\left(\mathbf{k}_{i}\right)=\omega_{s}\left(\mathbf{k}_{i}\right)=\frac{k_{i}v_{s}}{\left(1+k^{2}_{i}\lambda^{2}_{De}\right)^{1/2}},
\label{71}
\end{eqnarray}
\begin{eqnarray}
&\displaystyle
\gamma\left(\mathbf{k}_{i}\right)=\gamma_{e}\left(\mathbf{k}_{i}\right)-\gamma_{i}\left(\mathbf{k}_{i}\right),
\label{72}
\end{eqnarray}
where
\begin{eqnarray}
&\displaystyle
\gamma_{e}\left(\mathbf{k}_{i}\right)=\left(\frac{\pi}{8}\frac{m_{e}}{m_{i}}\right)^{1/2}
\frac{k_{i}v_{s}}{\left(1+k^{2}_{i}\lambda^{2}_{De}\right)^{2}}\left(\frac{k_{iz}V_{ez}}{\omega_{s}}-1\right),
\label{73}
\end{eqnarray}
and
\begin{eqnarray}
&\displaystyle
\gamma_{i}\left(\mathbf{k}_{i}\right)=\left(\frac{\pi}{8}\right)^{1/2}\frac{\omega^{4}_{s}}{k^{3}v^{3}_{Ti}}\exp\left(-\frac{\omega^{2}}{2k^{2}v^{2}_{Ti}}
\right).
\label{74}
\end{eqnarray}
It follows from Eqs. (\ref{72}) - (\ref{74}), that the ion-acoustic instability develops when $k_{iz}V_{ez}>\omega_{s}$. Neglecting the damping 
of the ion - acoustic waves on ions, determined by the decrement $\gamma_{i}\left(\mathbf{k}_{i}\right)$, the simple estimates\cite{Akhiezer1} follows from 
Eqs. (\ref{72}), (\ref{73}) for the frequency,
\begin{eqnarray}
&\displaystyle
\omega\left(\mathbf{k}_{i}\right)\sim \omega_{pi},
\label{75}
\end{eqnarray}
and the growth rate,
\begin{eqnarray}
&\displaystyle
\gamma\left(\mathbf{k}_{i}\right)\sim \omega_{pi}\frac{V_{ez}}{v_{Te}},
\label{76}
\end{eqnarray}
for the perturbations with $k_{i}\lambda_{De}\sim 1$.  For the data employed above in the analysis of the hydrodynamic current driven instabilities, 
we find for the argon plasma, that $V_{ez}$ determined by Eq. (\ref{27}) is equal to $10^{7}\, cm/sec$, $v_{s}= 2.6\cdot 10^{5}\, cm/sec < V_{ez}$, 
$\omega=\omega_{pi}=6.6\cdot 10^{7} \,sec^{-1}$, $v_{Te}=7\cdot10^{7}\, cm/sec$, and $\gamma=0.9\cdot10^{7}\,sec^{-1}$. Note, that during time 
$\gamma^{-1}$, electrons pass the distance only about $1.1\,cm$ with above estimated velocity $V_{ez}$. Therefore, it may be concluded, that the 
ion-acoustic instability has a possibility for the development in the bulk of plasma past the skin layer due to the electron current formed 
in the skin layer by the ponderomotive force. 

\section{Conclusions}\label{sec5}
In this paper, the stability theory of the inductively coupled plasma is developed for the high frequency regime, at which the classical skin 
effect develops. This theory is grounded on the methodology of the convective-oscillating modes, developed in Secs. \ref{sec2} and \ref{sec3}. It accounts 
for the oscillating motion of electrons in RF field jointly with their uniformly accelerated motion relative to ions under the action of the ponderomotive 
force. The theory reveals that the accelerated motion of electrons formed in the skin layer is the dominant 
factor in the temporal evolution of the electrostatic perturbed potential in the inductively coupled plasma and the current driven instabilities are more 
plausible for the spatially decaying oscillating RF field of the skin layer than the supposed parametric instabilities. 

We derive, that the current driven hydrodynamic instabilities, the dispersion properties of which are determined by Eq. (\ref{60}), which develop when the 
accelerated electron flow velocity $V_{ez}$ is much larger than the electron thermal velocity, do not meet the requirements for their development 
in the skin layer. We found, that the ion-acoustic current driven instability may exist in the inductively coupled plasmas under the action of the strong RF 
wave.  It develops with frequency (\ref{71}) and growth rate (\ref{72}) in the dipper regions of the bulk of plasma past the skin layer, where any 
restrictions specific for the skin layer are absent. In this regions, the electron flow formed in the skin layer moves with almost uniform velocity 
(\ref{27}), which is less than the electron 
thermal velocity, but is larger than the ion acoustic velocity. The development of this instability by RF wave will affect the absorption of the RF wave 
in plasma, the transition of the RF energy into the bulk of plasma and anomalous heating of the electrons and ions in the inner layers of the inductively 
coupled plasma. The scattering of the electrons by the ion-acoustic turbulence is the origin of the anomalous resistivity, which can be orders of magnitude 
larger than the resistivity due to electron-ion collisions\cite{Galeev, Bychenkov}, and of the resulted anomalous heating of the electrons and ions. 
The development of the ion-acoustic turbulence, powered by RF wave through the excitation of the ion-acoustic instability in the bulk of plasma is the 
basic mechanism of the nonlinear absorption of the RF weave energy in the regime of the normal skin effect. 
The theory of the ion-acoustic turbulence powered by the ion-acoustic current driven instability, which occurs in the bounded semi-infinite plasma under the 
action of the strong RF wave, was not developed yet. 

\begin{acknowledgments}
This work was supported by National R\&D Program through the National Research Foundation of 
Korea (NRF) funded by the Ministry of Education, Science and Technology (Grant No. NRF--2018R1D1A1B07050372) and BK21 Plus Creative 
Human Resource Development Program for IT Convergence.
\end{acknowledgments}

\bigskip
{\bf DATA AVAILABILITY}

\bigskip
The data that support the findings of this study are available from the corresponding author upon reasonable request.

\end{document}